\documentclass[10pt,a4paper]{article}
\usepackage{amsmath}
\usepackage{amsfonts}
\usepackage{amssymb}
\usepackage[dvips]{graphicx}
\usepackage{bm}

\newcommand{\be}{\begin{equation}}
\newcommand{\ee}{\end{equation}}
\newcommand{\ba}{\begin{array}}
\newcommand{\ea}{\end{array}}

\begin{document}

\title{\textbf{Internal structure of charged black holes}}
\author{\textsc{Dong-il Hwang}$^{a}$\footnote{dongil.j.hwang@gmail.com}\;  and \textsc{Dong-han Yeom}$^{b,c}$\footnote{innocent.yeom@gmail.com}\\
\textit{$^{a}$\small{Department of Physics, KAIST, Daejeon 305-701, Republic of Korea}}\\
\textit{$^{b}$\small{Center for Quantum Spacetime, Sogang University, Seoul 121-742, Republic of Korea}}\\
\textit{$^{c}$\small{Research Institute for Basic Science, Sogang University, Seoul 121-742, Republic of Korea}}}
\maketitle

\begin{abstract}
We investigate the internal structure of charged black holes with a spherically symmetric model in the double--null coordinate system.
Hawking radiation is considered using the S--wave approximation of semiclassical back--reaction and discharge is simulated by supplying an oppositely charged matter to the black hole.
In the stage of formation, the internal structure is determined by the mass and charge of collapsing matter.
When the charge--mass ratio is small, a wormhole--like internal structure is observed.
The structure becomes analogous to the static limit as the ratio reaches unity.
After the formation, mass inflation induces large curvature in the internal structure, which makes the structure insensitive to the late--time perturbations.
The internal structure determined from the formation seems to be maintained during a substantial mass reduction.
The discharge and neutralization of charged black holes is also investigated for both non--evaporating and evaporating cases.
Finally, we discuss the implications of the wormhole--like structure inside of charged black holes.
\end{abstract}


\newpage

\tableofcontents

\newpage

\section{\label{sec:Introduction}Introduction}
\begin{figure}
\begin{center}
\includegraphics[scale=0.5]{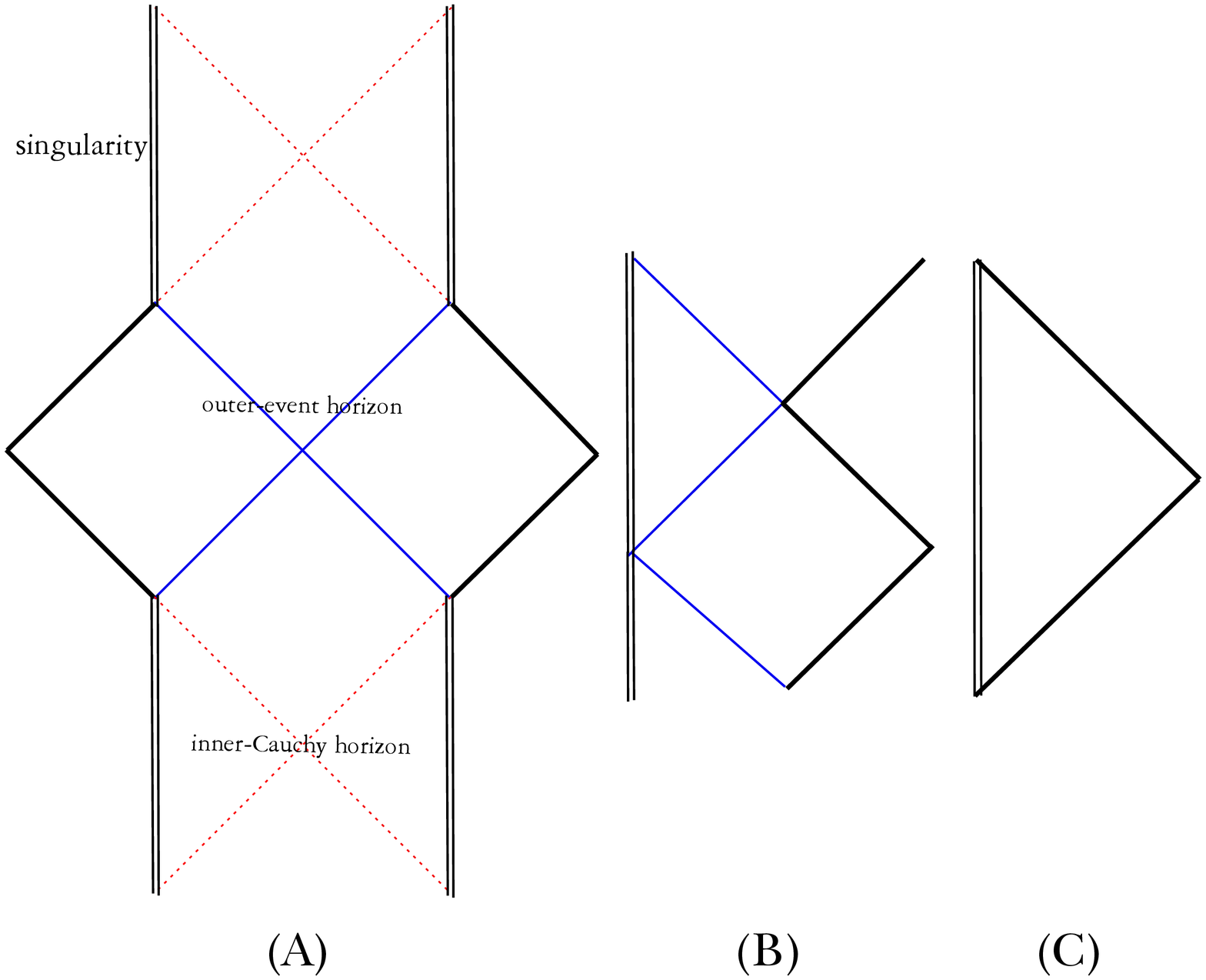}
\caption{\label{fig:charged_metric}Penrose diagrams of static charged black holes for (a) $M > Q$, (b) $M = Q$, and (c) $M < Q$.}
\end{center}
\end{figure}

\begin{figure}
\begin{center}
\includegraphics[scale=0.5]{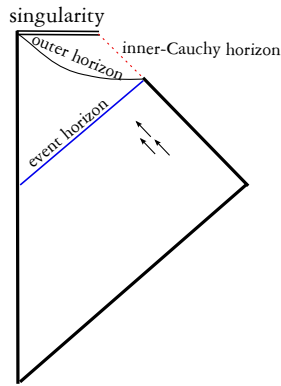}
\caption{\label{fig:charged_with_mass_inflation}Charged black holes with mass inflation scenario. The collapsing matter generates mass inflation which induces the space--like singularity, and the inner--Cauchy horizon becomes a null curvature singularity.}
\end{center}
\end{figure}

Charged black holes have complicated internal structures.
The Reissner-Nordstrom (RN) solution, a static electro-vacuum solution with spherical symmetry, is a good starting point from which to look over the structure of charged black holes \cite{RN} \cite{Hawking:1973uf}.
The RN solution has three different types of structure depending on the mass ($M$) and charge ($Q$) of the black hole (Figure \ref{fig:charged_metric}).
All of them have time--like singularities, and they induce Cauchy horizons which are boundaries of the causally undetermined regions \cite{Wald:1984rg}.

The existence of the time--like singularity and the Cauchy horizon has caused some issues related to cosmic censorship conjecture.
If $Q$ becomes greater than $M$, the naked singularity will violate weak cosmic censorship \cite{Vaidya}.
However, it is known that the electrostatic repulsion prevents such configurations \cite{OrenPiran}.
On the other hand, if an observer can pass through the Cauchy horizon, he or she will see the effects of the singularity violating strong cosmic censorship \cite{Wald:1984rg}.
However, in realistic charged black holes, the Cauchy horizon becomes a null curvature singularity via mass inflation \cite{Tipler}\cite{Burko_etal}\cite{Poisson:1990eh}\cite{Bonanno:1994ma}.
It plays a role as an impenetrable barrier to the observer, preserving strong cosmic censorship \cite{Burko_etal}\cite{Bonanno:1994ma}.
The back reaction of mass inflation also deforms the causal structure of charged black holes, as shown in Figure \ref{fig:charged_with_mass_inflation}.

As time passes, $M$ and $Q$ evolve via Hawking radiation and discharge, which can deform the structure of black holes.
When $M$ approaches to $Q$, a non-extreme charged black hole will approach to an extreme one.
On the other hand, when $Q$ is decreased to nearly zero, the charged black hole will become a neutral one.
In both cases, proper descriptions of the transition processes are required.

\begin{figure}
\begin{center}
\includegraphics[scale=0.5]{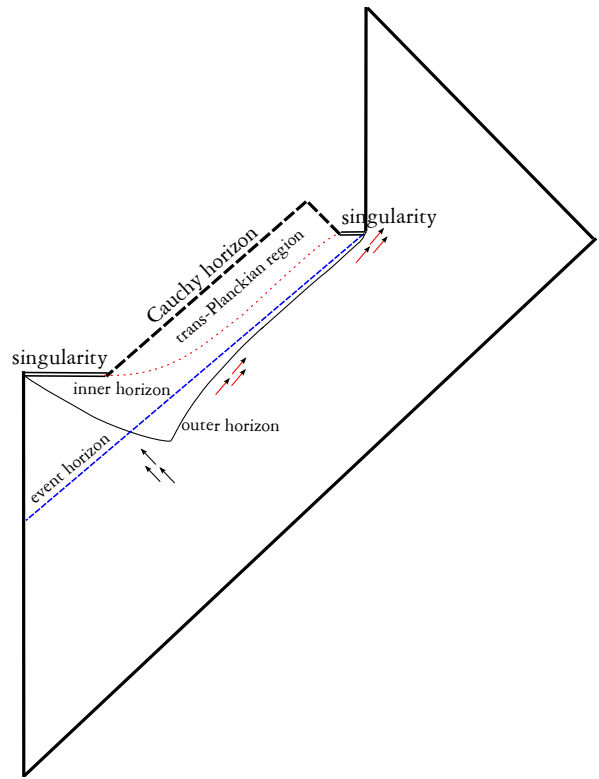}
\caption{\label{fig:charged_dynamic}Causal structure of dynamical charged black holes. Evaporation and neutralization are considered.}
\end{center}
\end{figure}

To figure out the dynamic process of charged black holes with quantum effects, some numerical approaches have been studied \cite{OrenPiran}\cite{Piran}\cite{HodPiran}\cite{SorkinPiran}\cite{Hansen:2005am}\cite{FKT}\cite{Hong:2008mw}.
When Hawking radiation was considered, the separation between the inner horizon and the Cauchy horizon was observed \cite{SorkinPiran}.
The inner horizon bends in a space--like direction and approaches to the time--like outer horizon.
This explains how a non-extreme charged black hole approaches to an extreme one where the two horizons coincide \cite{SorkinPiran}\cite{Hong:2008mw}.
On the other hand, as a charged black hole takes in some oppositely charged matter, its space--like inner horizon shrinks and becomes a singularity \cite{Hong:2008mw}.
This corresponds to the space--like singularity of a neutral black hole, and the Cauchy horizon from this singularity eliminates the internal structure.
This explains how the structure of charged black holes becomes a neutral structure.
According to the points mentioned above, we obtain the causal structure of charged black holes in Figure \ref{fig:charged_dynamic} \cite{Hong:2008mw}.

In this paper, we focus on the internal structure of charged black holes.
We examined the horizon structure inside of the outer horizon for important stages in time evolution: formation, evaporation, and neutralization of the black hole.
Therefore, the present work is also a study of the time evolution of charged black holes.
In Section \ref{sec:Model}, the numerical scheme is introduced.
In Section \ref{sec:BriefTimeEvolution}, we briefly review the time evolution of charged black holes.
In Section \ref{sec:InsideStructure}, the formation stage of charged black holes with various $M$ and $Q$ is studied.
In Section \ref{sec:ExoticMatter}, we investigate the response of charged black holes due to a substantial mass reduction using an exotic matter field.
In Section \ref{sec:Neutralization}, the neutralization process is discussed in detail.
In Section \ref{sec:Wormhole}, we explore the wormhole--like structure inside of charged black holes.
The convergence and consistency of the results are discussed in Appendix \ref{sec:Consistency}.

\section{\label{sec:Model}Model for a dynamical charged black hole}
\subsection{\label{sec:BasicSchemes}Equations and integration schemes}

We followed the numerical setup of the previous researches \cite{OrenPiran}\cite{Piran}\cite{HodPiran}\cite{SorkinPiran}\cite{Hansen:2005am}\cite{Hong:2008mw}\cite{Hansen:2009kn}\cite{Hwang:2010aj}.
In this paper, we set $G=c=k_{\mathrm{B}}=4\pi \epsilon_{0}=1$, and $m_{\mathrm{Pl}}=l_{\mathrm{Pl}}=\sqrt{\hbar}$, where $m_{\mathrm{Pl}}$ and $l_{\mathrm{Pl}}$ are the Planck mass and length, respectively.

We described a gauge--invariant Lagrangian for a complex massless scalar field $\phi$ coupled with an electromagnetic field $A_{a}$, and an exotic real scalar field $\psi$ \cite{Hawking:1973uf}
\begin{eqnarray} \label{Lagrangian}
\mathcal{L} = - \left(\phi_{;a} + i e A_{a} \phi) g^{ab} (\overline{\phi}_{;b} - i e A_{b} \overline{\phi}\right) + \psi_{;a} g^{ab} \psi_{;b} - \frac{1} {8 \pi} F_{ab} F^{ab},
\end{eqnarray}
where $F_{ab} = A_{b;a} - A_{a;b}$ and $e$ is the gauge coupling.
We also considered the semiclassical Einstein equation to include gravity and Hawking radiation,
\begin{eqnarray} \label{Einstein}
G_{ab} = 8\pi \left(T_{ab}^{C}+\langle \hat{T}_{ab}^{H} \rangle\right),
\end{eqnarray}
where $\langle \hat{T}_{ab}^{H} \rangle$ is the renormalized energy--momentum tensor.

Simply put, we make various charged black holes with $\phi$ and observe the influence of Hawking radiation and discharge.
A semiclassical back--reaction $\langle \hat{T}_{ab}^{H} \rangle$ is considered to model Hawking radiation.
We also included an exotic matter field $\psi$ which is a hypothetical material violating the energy condition.
It is used to reduce the mass of the black hole substantially in a dynamic way.
Discharge is simulated by supplying an oppositely charged scalar field to the black hole.
Detailed descriptions of each process are discussed in the following sections.

We use the double--null coordinate system assuming spherical symmetry
\begin{eqnarray} \label{DoubleNull}
ds^{2} = - \alpha^{2}(u,v) du dv + r^{2}(u,v) d\Omega^{2},
\end{eqnarray}
where $\alpha$ is the lapse function, $r$ is the radial function, and $u$ and $v$ indicate the in--going and out--going null directions, respectively \cite{Hamade:1995ce}.
Spherical symmetry allows a gauge choice where the gauge field is non-zero only in the $u$ component: $A_{a}=\left(a(u,v),0,0,0\right)$ \cite{OrenPiran}.

From this setup, the following components of the Einstein equation are calculated
\begin{eqnarray} \label{GAndT}
G_{uu} &=& -\frac{2}{r} \left(r_{,uu} - 2 \frac{\alpha_{,u}}{\alpha}r_{,u}\right),\\
G_{uv} &=& \frac{1}{2r^{2}} \left(4 r r_{,uv} + 4 r_{,u} r_{,v} + \alpha^{2}\right),\\
G_{vv} &=& -\frac{2}{r} \left(r_{,vv} - 2 \frac{\alpha_{,v}}{\alpha}r_{,v}\right),\\
G_{\theta \theta} &=& -4 \frac{r^2}{\alpha^{2}} \left( \frac{\alpha_{,uv}}{\alpha} + \frac{r_{,uv}}{r}-\frac{\alpha_{,u}\alpha_{,v}}{\alpha^2}\right),\\
T_{uu}^{C} &=& \phi_{,u} \overline{\phi}_{,u} + i e a\left(\phi \overline{\phi}_{,u} - \overline{\phi} \phi_{,u}\right) + e^{2} a^{2} \phi \overline{\phi} - \psi_{,u}^{2},\\
T_{uv}^{C} &=& \frac{{a_{,v}}^{2}}{4\pi\alpha^{2}},\\
T_{vv}^{C} &=& \phi_{,v} \overline{\phi}_{,v}  - \psi_{,v}^{2},\\
T_{\theta\theta}^{C} &=& \frac{r^{2}}{\alpha^{2}} \left(\phi_{,u} \overline{\phi}_{,v} + \overline{\phi}_{,u} \phi_{,v} + i e a\left(\phi \overline{\phi}_{,v} - \overline{\phi} \phi_{,v}\right) - 2 \psi_{,u} \psi_{,v}+ \frac{{a_{,v}}^{2}}{2\pi \alpha^{2}}\right).
\end{eqnarray}

Equations for the scalar field and the electromagnetic field are
\begin{eqnarray} \label{ScalarAndMaxwell2}
r \phi_{,uv} + r_{,u} \phi_{,v} + r_{,v} \phi_{,u} + i e a r \phi_{,v} + i e a r_{,v} \phi + \frac{i}{2}{e a_{,v} r \phi} &=& 0,\\
r \psi_{,uv} + r_{,u} \psi_{,v} + r_{,v} \psi_{,u} &=& 0,\\
\left( \frac{r^{2} a_{,v}}{\alpha^{2}} \right)_{,v} - i \pi r^{2} e \left(\phi \overline{\phi}_{,v} - \overline{\phi} \phi_{,v}\right) &=& 0.
\end{eqnarray}
We define $q \equiv 2r^{2}a_{,v}/\alpha^2$, which can be interpreted as the electric charge within a sphere of radius $r$.

We include the semiclassical back--reaction in one--loop order from the renormalized energy-momentum tensor.
Because of the spherical symmetry, we use the $(1+1)$--dimensional result divided by $4\pi r^{2}$ as an S--wave approximation \cite{SorkinPiran}\cite{Davies:1976ei}
\begin{eqnarray} \label{semiclassical}
\langle \hat{T}_{uu}^{H} \rangle &=& \frac{P}{4\pi r^{2}}\left(\frac{\alpha_{,uu}}{\alpha} - 2 \frac{\alpha_{,u}^{2}}{\alpha^{2}}\right),\\
\langle \hat{T}_{uv}^{H} \rangle = \langle \hat{T}_{vu}^{H} \rangle &=& -\frac{P}{4\pi r^{2}}\left(\frac{\alpha_{,uv}}{\alpha} - \frac{\alpha_{,u} \alpha_{,v}}{\alpha^{2}}\right),\\
\langle \hat{T}_{vv}^{H} \rangle &=& \frac{P}{4\pi r^{2}}\left(\frac{\alpha_{,vv}}{\alpha} - 2 \frac{\alpha_{,v}^{2}}{\alpha^{2}}\right),
\end{eqnarray}
where $P \equiv Nl_{\mathrm{Pl}}^2 / 12\pi$, and $N$ is the number of massless scalar fields generating Hawking radiation.
$P$ determines the strength of the Hawking radiation.
For given $P$, $N$ determines the trans--Planckian scale where the semiclassical approximation breaks down \cite{Hong:2008mw}.
Therefore, as we assume larger $N$, our approximation becomes more reliable.
On the other hand, the expectation value of quantum operators has physical meanings on a classical background only when the quantum dispersion is relatively small.
Large $N$ assumption also satisfies this condition because each field independently contributes to the energy-momentum tensor, resulting in a small total dispersion \cite{Hong:2008mw}.
Throughout this paper, we assume that $N$ is sufficiently large.

For numerical integrations, we set equations following the conventions of previous researches \cite{OrenPiran}\cite{HodPiran}\cite{SorkinPiran}
\begin{eqnarray} \label{Conventions}
&&d \equiv \frac{\alpha_{,v}}{\alpha},\;\;\; h \equiv \frac{\alpha_{,u}}{\alpha},\;\;\; f \equiv r_{,u},\;\;\; g \equiv r_{,v}, \nonumber \\
&&s \equiv \sqrt{4\pi} \phi,\;\;\; w \equiv s_{,u},\;\;\; z \equiv s_{,v}, \\
&&b \equiv \sqrt{4\pi} \psi,\;\;\; o \equiv b_{,u},\;\;\; p \equiv b_{,v}. \nonumber
\end{eqnarray}

\begin{enumerate}
\item \textit{Einstein equations}
\begin{eqnarray} \label{E1}
h_{,v} = d_{,u} &=& \frac{1}{\left(1-P/r^{2}\right)} \left( \frac{f g}{r^{2}} + \frac{\alpha^2}{4 r^{2}} - \frac{\alpha^{2} q^{2}}{2 r^{4}}
- \frac{1}{2} \left(\left(w\overline{z}+\overline{w}z\right) - i e a\left(s\overline{z}-\overline{s}z\right)\right) + o p\right),\\
\label{E2}
f_{,u} &=& 2fh - r \left(w\overline{w} + i e a\left(\overline{w}s-w\overline{s}\right) + e^{2}a^{2} s\overline{s}\right) + r o^{2} - \frac{P}{r}\left(h_{,u}-h^{2}\right),\\
\label{E3}
f_{,v} = g_{,u} &=& -\frac{f g}{r} - \frac{\alpha^{2}}{4r} + \frac{\alpha^{2} q^{2}}{4 r^{3}} - \frac{P}{r} d_{,u},\\
\label{E4}
g_{,v} &=& 2dg - r z\overline{z} + r p^{2} - \frac{P}{r}(d_{,v}-d^{2}).
\end{eqnarray}

\item \textit{Maxwell equations}
\begin{eqnarray} \label{M1}
a_{,v} &=& \frac{\alpha ^{2} q}{2 r^{2}},\\
\label{M2}
q_{,v} &=& -\frac{i e r^{2}}{2} \left(\overline{s}z-s\overline{z}\right).
\end{eqnarray}

\item \textit{Scalar field equations}
\begin{eqnarray} \label{S1}
w_{,v} &=& z_{,u} = - \frac{f z}{r} - \frac{g w}{r} - \frac{i e a r z}{r} - \frac{i e a g s}{r} - \frac{i e}{4r^{2}}\alpha^{2}q s,\\
\label{S3}
o_{,v} &=& p_{,u} = - \frac{f p}{r} - \frac{g o}{r}.
\end{eqnarray}
\end{enumerate}

We solve this set of equations using the second order Runge--Kutta method \cite{nr}.
$h_{,u}$ and $d_{,v}$ in Equations (\ref{E2}) and (\ref{E4}) are estimated by the finite difference method.
There are three equations evolving the radial function $r$: Equations (\ref{E2}), (\ref{E3}), and (\ref{E4}).
Integrations using any of them should yield the same results.
We choose Equation (\ref{E4}) in most of the simulations and use the others to check the consistency of the code (see Appendix \ref{sec:Consistency}).

Note that Equation (\ref{E1}) is singular at $r=\sqrt{P}$, which originates from the semiclassical approximation \cite{Piran}.
Therefore, for a reliable approximation, $\sqrt{P}$ should be sufficiently smaller than the typical length scale of the simulation.
This constraint is satisfied throughout this paper, and, when $P\neq0$, \emph{we regard $r=\sqrt{P}$ as the central singularity.}

Since we are working in the double--null coordinate system, $r_{,v} = 0$ and $r_{,u} = 0$ contours can be interpreted as the trapping and anti--trapping horizons, respectively.
The electric charge $q$ is defined above, and we use the Misner--Sharp mass function $m(u,v) = (r/2) (1+q^{2}/r^{2}+4r_{,u}r_{,v}/\alpha^{2})$ \cite{Waugh:1986jh}.
Therefore, $M$ and $Q$ correspond to $m$ and $q$ measured at the outer horizon after the black hole is formed and stabilized.

\subsection{\label{sec:InitialCond}Initial conditions}

We simulate the gravitational collapse of a charged matter shell; we place a pulse of a complex scalar field along an out--going null hypersurface, and collapse it to form a black hole.
We choose $u=\mathrm{const}$ surfaces as out--going null hypersurfaces and $v=\mathrm{const}$ surfaces as in--going null hypersurfaces.
Therefore, $u$ and $v$ can be interpreted as the retarded and advanced times, respectively.
We assign initial conditions for each function on $u=0$ and $v=0$ surfaces.

There is a gauge freedom in choosing the initial $r$ function.
We choose $r(u,0)=ur^{(0)}_{,u}+r^{(0)}$ and $r(0,v)=vr^{(0)}_{,v}+r^{(0)}$; therefore, $g(0,v)=r^{(0)}_{,v}$ and $f(u,0)=r^{(0)}_{,u}$.
Because the initial condition describes a shell--shaped scalar field, its inside is not affected by the shell.
Therefore, we can simply choose $q(u,0)=0$, $a(u,0)=0$, and $\alpha(u,0)=1$.
Since the mass function should vanish in this region, we choose $r^{(0)}_{,u}=-1/2$, $r^{(0)}_{,v}=1/2$ and set $r^{(0)}=r(0,0)=10$.

We place a complex scalar field $\phi$ along the $u=0$ surface,
\begin{eqnarray} \label{s1_initial}
\phi(0,v) = \left\{ \begin{array}{ll}
\frac{A_{1}}{\sqrt{4\pi}} \sin ^{2} \left( \pi \frac{v}{20} \right)\left(\cos \left(\pi \frac{v}{10} \right)
+ i \cos\left(\pi \frac{v}{10} + \delta \right)\right) & 0 \leq v < 20,\\
\frac{A_{2}}{\sqrt{4\pi}} \sin ^{2} \left( \pi \frac{v-20}{100} \right)\left(\cos \left(\pi \frac{v-20}{50} \right)
+ i \cos\left(\pi \frac{v-20}{50} - \frac{\pi}{2} \right)\right) & 20 \leq v < 120,\\
0 & 120 \leq v.\\
\end{array} \right.
\end{eqnarray}
In $0 \leq v < 20$, a pulse with an amplitude of $A_{1}/\sqrt{4\pi}$ collapses to form a black hole.
Its phase difference $\delta$ determines the amount of initial charge (see Equation (\ref{M2})); field configurations with $\delta=0$ have no charge, and the charge increases as $\delta$ increases until $\delta=\pi/2$.
When the discharge of the black hole is considered, we supply an oppositely charged pulse to the black hole, with an amplitude $A_{2}/\sqrt{4\pi}$ in $20 \leq v < 120$.

When the mass reduction via an exotic matter field is considered, we place $\psi$ along the $u=0$ surface,
\begin{eqnarray}
\psi(0,0) &=& 0,\\
\psi_{,v}(0,v) &=& \left\{ \begin{array}{ll}
0 & 0 \leq v < 20,\\
\frac{A_{3}}{\sqrt{4\pi}} \sin\left(\pi \frac{v-20}{20} \right) & 20 \leq v < 30,\\
\frac{A_{3}}{\sqrt{4\pi}} & 30 \leq v.
\end{array} \right.
\end{eqnarray}
The $T_{vv}^{C}$ component of exotic matter field smoothly decreases from zero to $- A_{3}^{2}/(4\pi)$ in $20 \leq v < 30$, and is uniformly distributed in $30 \leq v$.

We place no field along the $v=0$ surface.
Finally, we derive $d(0,v)=r\left(z\overline{z} - p^{2}\right)/(2g)$ from Equation (\ref{E4}) assuming that the Hawking effect is negligible on the initial surfaces \cite{SorkinPiran}.

We choose the domain of computation $0 \leq u \leq 20$ and $0 \leq v \leq 100$.
For all simulations, $e=0.3$ and $A_{1}=0.25$ are used to form an initial black hole.

\section{\label{sec:BriefTimeEvolution}Brief review of the time evolution of charged black holes}
\begin{figure}
\begin{center}
\includegraphics[scale=0.7]{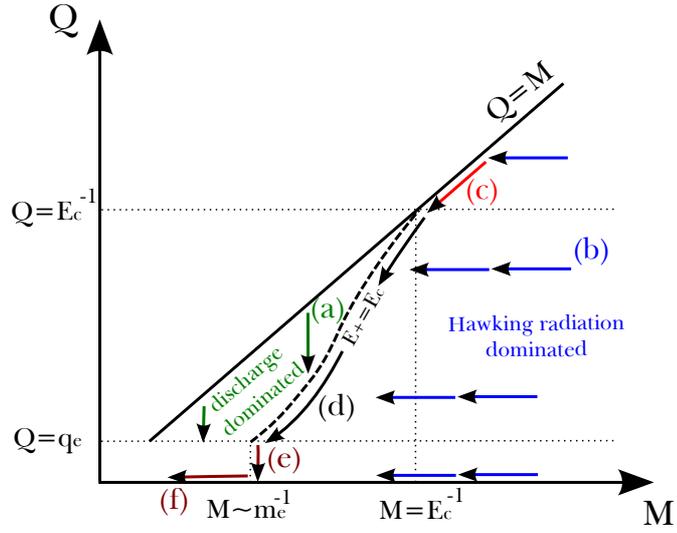}
\caption{\label{fig:MQevolution}Evolution of charged black holes \cite{Hong:2008mw}. (a) If $E_{+}\gg E_{c}$, discharge dominates and the black hole approaches $E_{+} = E_{c}$. (b) If $E_{+}\ll E_{c}$, discharge is exponentially suppressed. If $Q/M \ll 1$, Hawking radiation dominates and $Q/M$ decreases. (c) If $Q \gg E^{-1}_{c}$, the black hole will approach and follow the extreme state. (d) If $Q \ll E^{-1}_{c}$, the black hole will approach and follow $E_{+} = E_{c}$. (e) When the charge reduces to a few quanta, $Q \sim q_{e}$ and $T_{H}\sim M^{-1}\sim m_{e}$ on the $E_{+}=E_{c}$ track. Then, the black hole will emit its final quanta of charge via Hawking radiation. (f) The final neutral black hole will lose mass via Hawking radiation.}
\end{center}
\end{figure}

This section summarizes the time evolution of charged black holes based on previous researches \cite{SorkinPiran}\cite{Hong:2008mw}.
There are two quantum effects that determine the time evolution: Hawking radiation and discharge.
Hawking radiation is a thermal radiation of black holes, and $M$ is reduced as the energy is emitted.
It has Hawking temperature defined as:
\begin{eqnarray} \label{Temperature}
T_{H}=\frac{\sqrt{M^2-Q^2}}{2\pi r_{+}},
\end{eqnarray}
where $r_{+}=M + \sqrt{M^2-Q^2}$ is the radius of the outer horizon.
The use of Misner--Sharp mass function guarantees that $r_{+}$ is equivalent to the radius of outer most trapped surface defined by $r_{,v} = 0$.
As the black hole becomes extreme, its temperature becomes zero and the radiation stops.
Therefore, if one ignores discharge, the extreme charged black hole can be regarded as an eternal remnant.

When the electric field near the outer horizon is strong enough, Schwinger pair--creation in this region reduces $Q$ while slightly increasing $M$.
In this regime, the pair-creation rate can be approximated as:
\begin{eqnarray}
\Gamma_{+} \cong \frac{q_{e}^{2} E^{2}}{4 \pi^{3}} e^{-\frac{E_{c}}{E}},
\end{eqnarray}
where $E_{c}=\pi m_{e}^{2}/q_{e}$ is the critical pair--creating field for particles with mass $m_{e}$ and charge $q_{e}$ and $E_{+} = Q/r_{+}^{2}$ is the electric field near the outer horizon \cite{Dunne:2004nc}.
Note that for a given $Q/M$, the pair--creation is exponentially suppressed when the black hole is large.

The brief paths of time evolution, considering Hawking radiation and discharge, are described in Figure \ref{fig:MQevolution}, which is excerpted from the previous paper \cite{Hong:2008mw}.
For a sufficiently large black hole, pair creation is highly suppressed and $Q/M$ increases via Hawking radiation.
As $M$ becomes comparable to $Q$, Hawking temperature decreases and the black hole maintains a nearly extreme state, losing its mass and charge simultaneously.
After $Q$ becomes sufficiently small, discharge dominates Hawking radiation and $Q/M$ decreases.
It keeps decreasing until the black hole emits its final quanta of charge and becomes a neutral black hole \cite{Gibbons:1975kk}.
Finally, it will be totally evaporated, which is in the realm of quantum gravity.

\section{\label{sec:InsideStructure}Formations of charged black holes with various $Q/M$}
Charged black holes go through states with various $M$ and $Q$ in their time evolution.
In this section, we simulate gravitational collapses of matter shells with various masses and charges, considering Hawking radiation.
The internal structures of the resulting black holes are studied, which can help in understanding the time evolution.

\begin{figure}
\begin{center}
\includegraphics[scale=0.3]{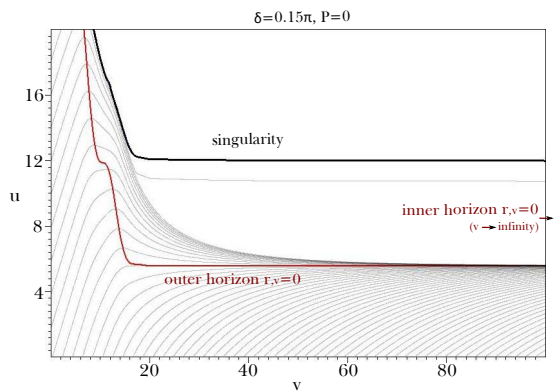}
\caption{\label{fig:d0.15P0r}Contour diagram of radial function $r$ and horizon structure for $\delta=0.15\pi$ and $P=0$. This diagram corresponds to a dynamically formed charged black hole within the static limit.}
\end{center}
\end{figure}

To see the net effect of Hawking radiation, we review the static limit of charged black holes first \cite{OrenPiran}\cite{HodPiran}.
Figure \ref{fig:d0.15P0r} shows the result with $P=0$, which is consistent with Figure \ref{fig:charged_with_mass_inflation}.
In a neutral black hole, the singularity and the outer horizon become stuck together in the double null coordinate system (see Figure \ref{fig:d0r}); that is, any time--like observers beyond the outer horizon inevitably hit the central singularity.
However, a gap arises between the singularity and the outer horizon in Figure \ref{fig:d0.15P0r}.
In this region, the radial function monotonically decreases to some non--zero values along the out--going null direction \cite{OrenPiran}, which indicates the existence of the inner--Cauchy horizon at $v \rightarrow \infty$.
The curvature and the mass function increase exponentially in this region, which is a clear sign of mass inflation \cite{Poisson:1990eh}.
It makes the inner--Cauchy horizon a null curvature singularity, which preserves strong cosmic censorship.
The qualitative behavior of charged black holes within the static limit does not vary much as we change $Q/M$.

We have two free parameters that determine the amount of initial charge: $e$ and $\delta$.
Since we are considering the time evolution, changing a physical constant $e$ is irrelevant.
Therefore, we change $\delta$ to control the initial charge and fix the other parameters.
We set an initial condition that makes a nearly extreme black hole when $\delta=\pi/2$, and repeat the simulation with slightly decreased $\delta$ until $\delta=0$ and the black hole becomes a neutral one.
We set $P=0.01$ to include Hawking radiation, and $A_{2}$ and $A_{3}$ are set to zero.
Among the results, we investigate some of typical ones in detail.

\subsection{\label{sec:d0}$\delta=0$ ($Q/M = 0$)}
\begin{figure}
\begin{center}
\includegraphics[scale=0.3]{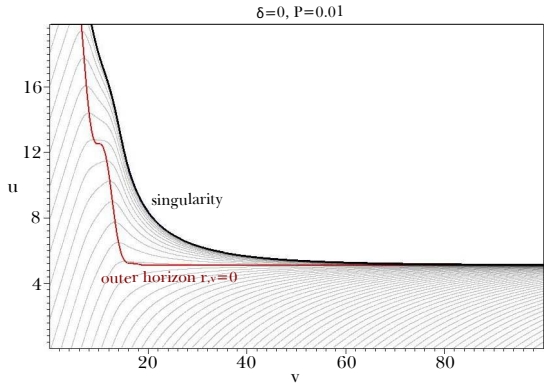}
\caption{\label{fig:d0r}Contour diagram of radial function $r$ and horizon structure for $\delta=0$. This diagram corresponds to an evaporating neutral black hole.}
\end{center}
\end{figure}

We first discuss the $\delta=0$ ($Q/M = 0$) case which corresponds to an evaporating neutral black hole.
The results are described in Figure \ref{fig:d0r}.
As the matter shell collapses in $0\leq v\leq 20$, the outer horizon grows in a space--like direction.
Then, by Hawking radiation, it bends in a time--like direction \cite{SorkinPiran}\cite{Hong:2008mw}.
Beyond the outer horizon, there is a space--like singularity where the curvature diverges.
These results are consistent with the well--known characteristics of neutral black holes \cite{inforpara}\cite{CGHS}\cite{local_horizon}.

\subsection{\label{sec:d0.05pi}$\delta=0.05\pi$ ($Q/M \simeq 0.23$)}
\begin{figure}
\begin{center}
\includegraphics[scale=0.3]{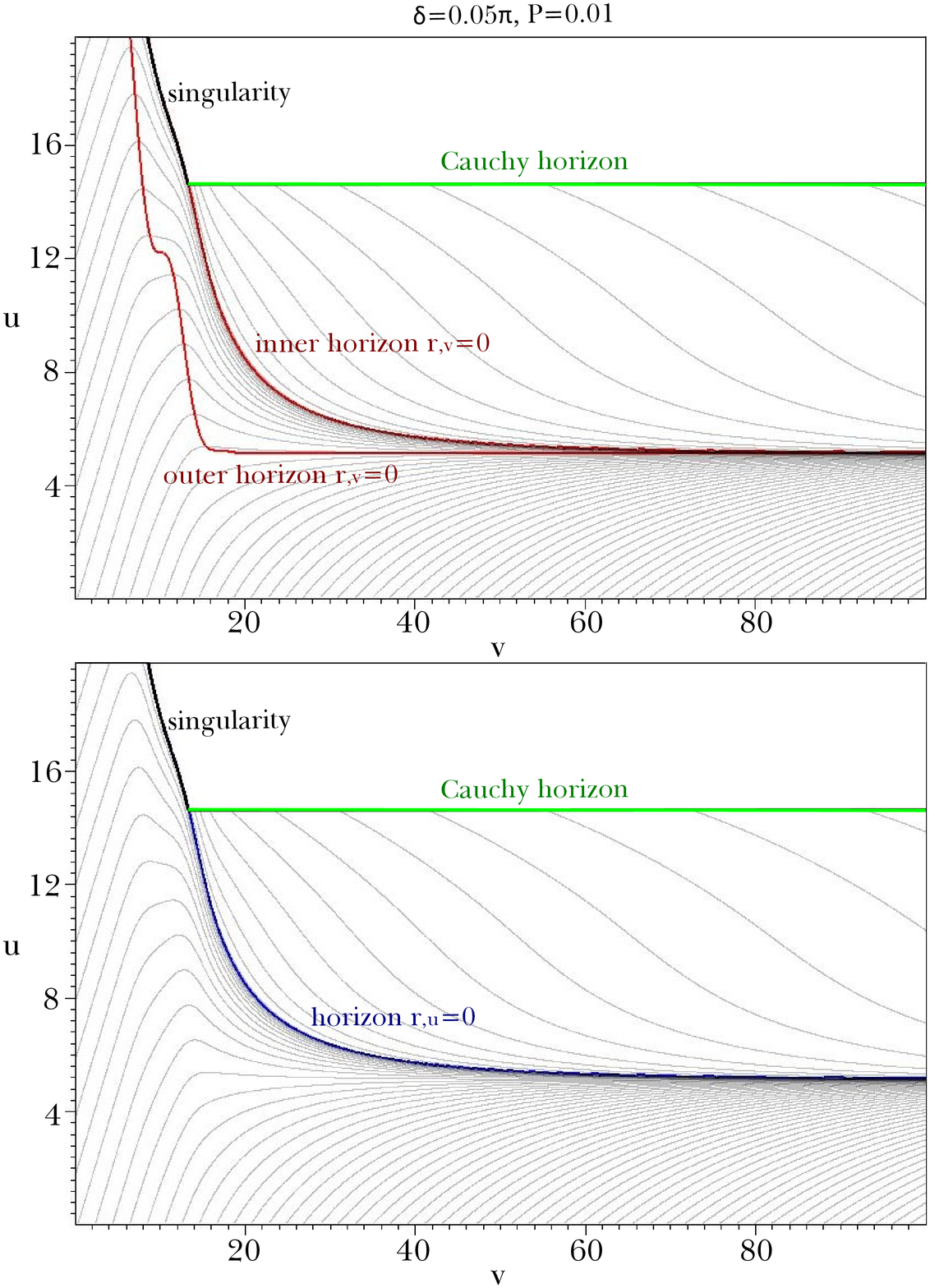}
\caption{\label{fig:d0.05r}Contour diagrams of radial function $r$ and horizon structures for $\delta=0.05\pi$. As we include Hawking radiation, both $r_{,u} = 0$ and $r_{,v} = 0$ types of inner horizons are observed. The resulting structure is similar to a throat of a wormhole.}
\end{center}
\end{figure}

With $\delta=0.05\pi$ ($Q/M \simeq 0.23$), the amount of charge is enough to generate the typical structures of charged black holes \cite{SorkinPiran}\cite{Hong:2008mw}.
In Figure \ref{fig:d0.05r}, $r_{,u} = 0$ and $r_{,v} = 0$ contours exist beyond the outer horizon that correspond to two different types of inner horizons.
When Hawking radiation is not considered, an out--going null observer escaping from a charged black hole possibly hits the null curvature singularity at the inner--Cauchy horizon and then falls into the central singularity \cite{Bonanno:1994ma}.
However, in an evaporating charged black hole, the observer reaches the $r_{,v} = 0$ type inner horizon, where the radial function starts to increase.
The $r_{,v} > 0$ region beyond this horizon causes a Cauchy horizon along the out--going null direction.
Note that, as discussed in Section \ref{sec:Introduction}, the curvature deep inside of this region becomes trans--Planckian because of mass inflation.
Nevertheless, the horizon structure and the region around the inner horizon are still reliable in a background of realizable $N$.
The finite curvature near the Cauchy horizon also has a possibility of violating strong cosmic censorship \cite{Hong:2008mw}.
These regular horizons make the causal structure of charged black holes similar to that of nonsingular black holes \cite{Hayward:2005gi}.

One interesting feature is the other type of inner horizon, $r_{,u} = 0$.
Both in a neutral black hole and a charged black hole within the static limit, an in--going null observer directly hits the central singularity.
However, in an evaporating charged black hole, an $r_{,u} > 0$ region exists beyond the $r_{,u} = 0$ type inner horizon.
Since this region is inside of the $r_{,v} = 0$ type inner horizon, the radial function or, equivalently, the locally measured area increases for any physical observers.
Therefore, as we include Hawking radiation, the internal structure of charged black holes becomes analogous to a throat of a wormhole.
This topic is discussed in Section \ref{sec:Wormhole}.

\subsection{\label{sec:d0.15pi}$\delta=0.15\pi$ ($Q/M \simeq 0.61$)}
\begin{figure}
\begin{center}
\includegraphics[scale=0.3]{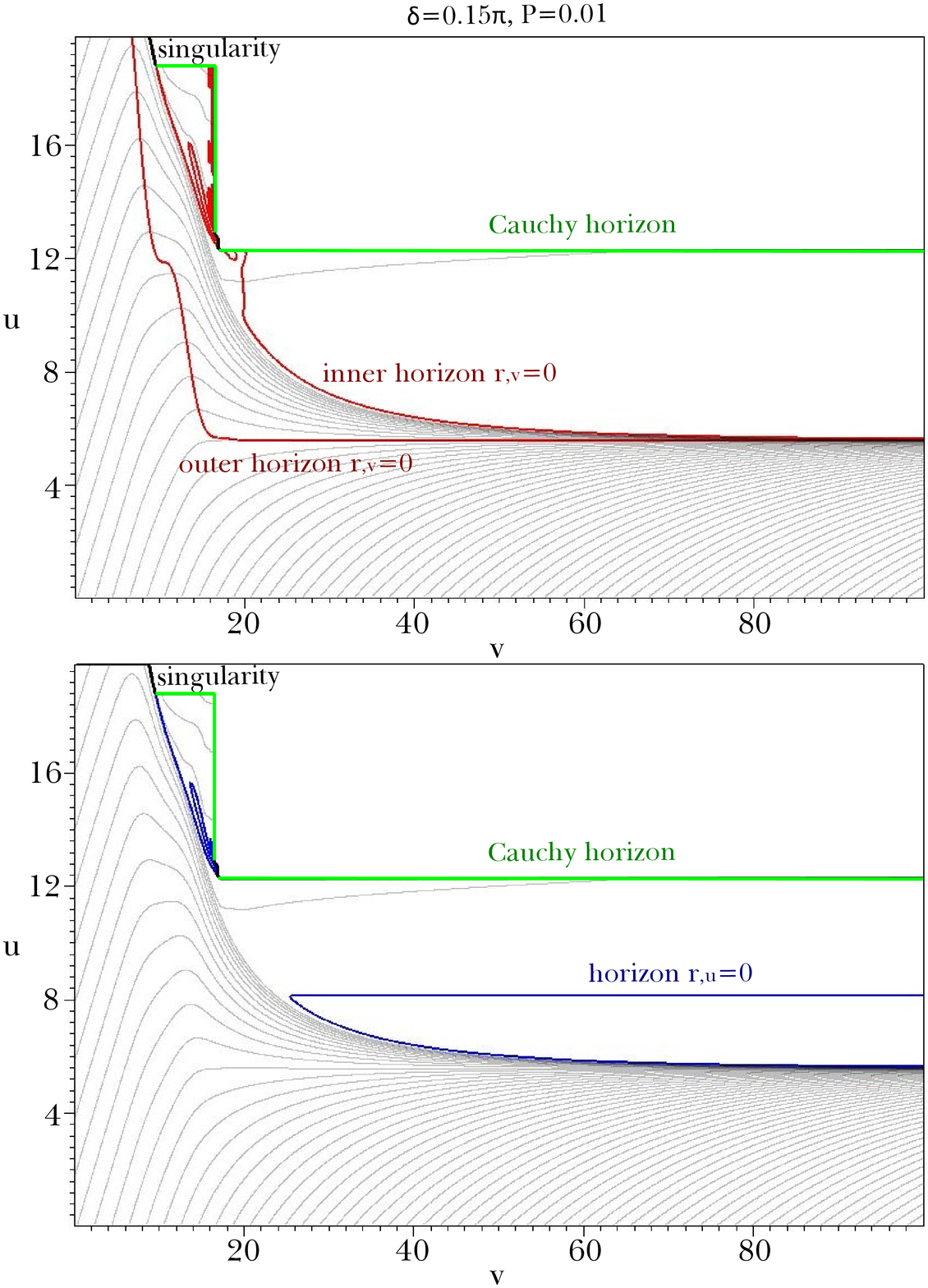}
\caption{\label{fig:d0.15r}Contour diagrams of radial function $r$ and horizon structures for $\delta=0.15\pi$. As the charge is increased, the $r_{,u} > 0$ region inside of the black hole is decreased. On the other hand, the qualitative behavior of the $r_{,v} = 0$ type inner horizon is unchanged.}
\end{center}
\end{figure}

\begin{figure}
\begin{center}
\includegraphics[scale=0.7]{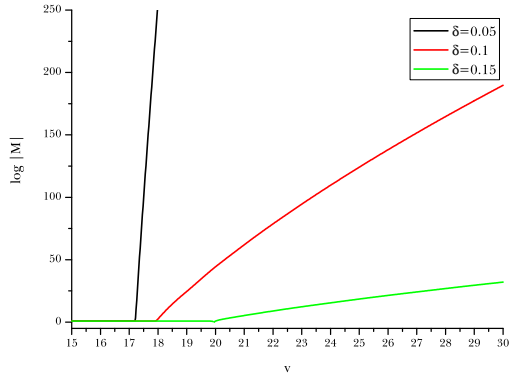}
\caption{\label{fig:massinfvsdelta}Plot of $\log |M|$ along $u=10$ surface for black holes with $\delta=0.05\pi$, $\delta_{1}=0.1\pi$, and $\delta_{1}=0.15\pi$. This plot clearly shows the exponential behavior of the mass inflation effect, which becomes stronger as $Q/M$ decreases.}
\end{center}
\end{figure}

We investigate a black hole with more charge, $\delta=0.15\pi$ ($Q/M \simeq 0.61$), in Figure \ref{fig:d0.15r}.
The qualitative behavior of the outer horizon and the $r_{,v} = 0$ type inner horizon is the same as in Figure \ref{fig:d0.05r}.
On the other hand, the $r_{,u} = 0$ type inner horizon structure is changed from Figure \ref{fig:d0.05r}.
A boundary arises along an out--going null direction, where $r_{,u}$ becomes negative again.
This boundary moves downward as $Q/M$ increases, and when $Q/M \sim 1$, the $r_{,u} > 0$ region is not observed.

The mass inflation effect also depends on $Q/M$ of black holes.
In the RN metric, mass inflation affects the mass function according to $m(v) \sim \exp(\kappa_{\mathrm{i}}v)$, where $\kappa_{\mathrm{i}}$ is the surface gravity at the inner horizon which increases as $Q/M$ decreases \cite{Poisson:1990eh}.
We plot the observed $\log |M|$ in Figure \ref{fig:massinfvsdelta}, and it clearly shows the exponential behavior of mass function \cite{OrenPiran}\cite{HodPiran}.
Its exponent increases as $Q/M$ decreases, which is consistent with the calculation in the RN metric.

\subsection{\label{sec:d0.3pi}$\delta=0.3\pi$ ($Q/M \simeq 0.91$)}
\begin{figure}
\begin{center}
\includegraphics[scale=0.3]{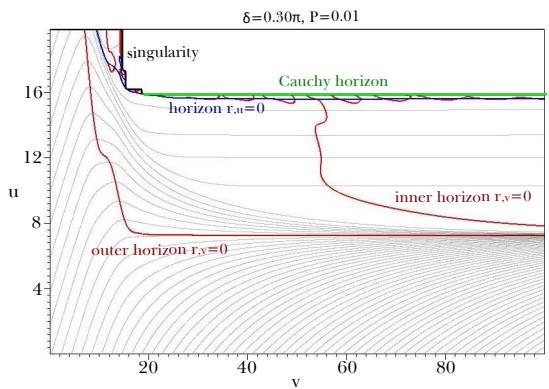}
\caption{\label{fig:d0.30r}Contour diagram of radial function $r$ and horizon structure for $\delta=0.3\pi$. Among massy horizons, there exists a distinct $r_{,v} = 0$ type inner horizon. The $r_{,u} > 0$ region becomes negligible.}
\end{center}
\end{figure}

Figure \ref{fig:d0.30r} describes a charged black hole with $\delta=0.3\pi$ ($Q/M \simeq 0.91$).
Beyond the outer horizon, the contour lines of radial function become quite flat along the out--going null direction.
It makes small fluctuations in $r$ contour form complicated $r_{,v} = 0$ horizons, which do not seem to be very meaningful.
However, among the messy horizons, a distinct $r_{,v} = 0$ type inner horizon exists which approaches to the outer horizon.

\subsection{\label{sec:d0.5pi}$\delta=0.5\pi$ ($Q/M \simeq 0.98$)}
\begin{figure}
\begin{center}
\includegraphics[scale=0.3]{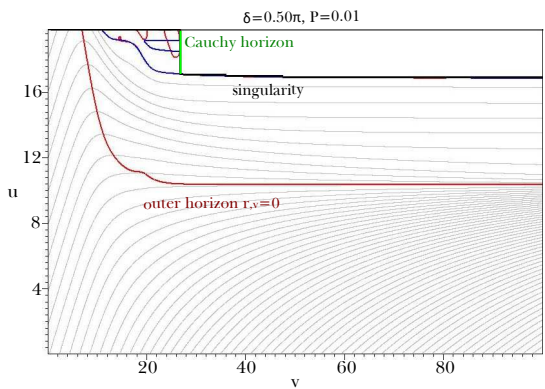}
\caption{\label{fig:d0.50r}Contour diagram of radial function $r$ and horizon structure for $\delta=0.5\pi$. The black hole is nearly extreme. Its structure is analogous to the static limit in Figure \ref{fig:d0.15P0r}.}
\end{center}
\end{figure}

Now we investigate a nearly extreme black hole, $\delta=0.5\pi$ ($Q/M \simeq 0.98$), in Figure \ref{fig:d0.50r}.
It has very low Hawking temperature, which implies that the semiclassical back--reaction is negligible.
One can reasonably expect that its structure should be analogous to the static limit in Figure \ref{fig:d0.15P0r}.
Interestingly, their horizon structures are basically the same even though their mass and charge are quite different.
Neither of them have inner horizons nor Cauchy horizons in the scope of simulation.
This observation also supports the physical reliability of the model.

\subsection{\label{sec:4summary}Summary}
\begin{figure}
\begin{center}
\includegraphics[scale=0.6]{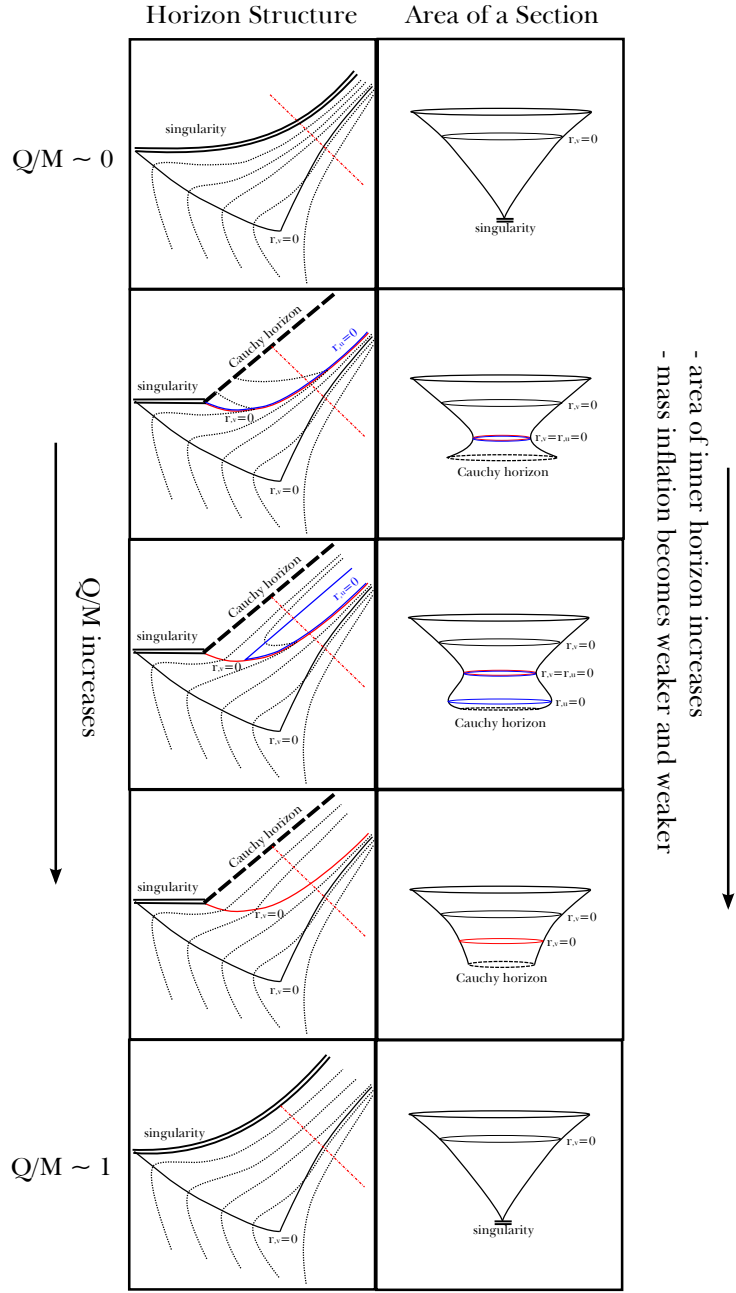}
\caption{\label{fig:diagram_new}Formation of charged black holes with various $Q/M$. Diagrams in the left column show the horizon structures and the right column show schematic diagrams of the area measured by an in--going null observer along the red dotted line in the corresponding left diagram.}
\end{center}
\end{figure}

The results in this section are summarized in Figure \ref{fig:diagram_new}.
As we consider black holes with sufficient charge, complicated internal structures arise.
Black holes with relatively small $Q/M$ have both types of inner horizons, and the resulting structure is similar to a throat of a wormhole.
In black holes with larger $Q/M$, the $r_{,u} > 0$ region gradually disappears, and eventually, neither types of inner horizons are observed in nearly extreme black holes.

It was known that Hawking radiation significantly deforms the internal structure of charged black holes for relatively small $Q/M$ \cite{SorkinPiran}\cite{Hong:2008mw}.
On the other hand, Hawking radiation should be negligible for nearly extreme black holes.
The observations in this section smoothly connect these two claims.
Note that it cannot be applied directly to the time evolution of internal structure.
This topic is discussed in the following section.

\section{\label{sec:ExoticMatter}Mass reduction via an exotic matter field}
The semiclassical back--reaction approximates Hawking radiation in a simple and proper way.
Nevertheless, it has some drawbacks in simulating the long--term time evolution.
First of all, it has an artificial singularity at $r=\sqrt{P}$.
Since $\sqrt{P}$ should be sufficiently smaller than the typical length scale of the simulation, it imposes an upper bound on the value of $P$.
Therefore, if one wants to observe a substantial mass evaporation, one should carry out a long--term simulation rather than using large $P$ for strong Hawking radiation.
However, there is a technical difficulty that the radial contours badly fold around the asymptotic horizons in long--term simulations \cite{OrenPiran}.
To observe the substantial mass evaporation, another approximation scheme is needed.

In this section, we supply an exotic matter field to the black hole and study how the internal structures are affected by the substantial mass reduction.
This observation is important because black holes with the same mass and charge can have various internal structures depending on their different histories.
By simulating a dynamic process of mass reduction, we can get some hints on how internal structures will be affected during the long--term evaporation.
For all simulations, we set $A_{2}=0$, $\delta=0.15\pi$, and $P=0$.

\begin{figure}
\begin{center}
\includegraphics[scale=0.3]{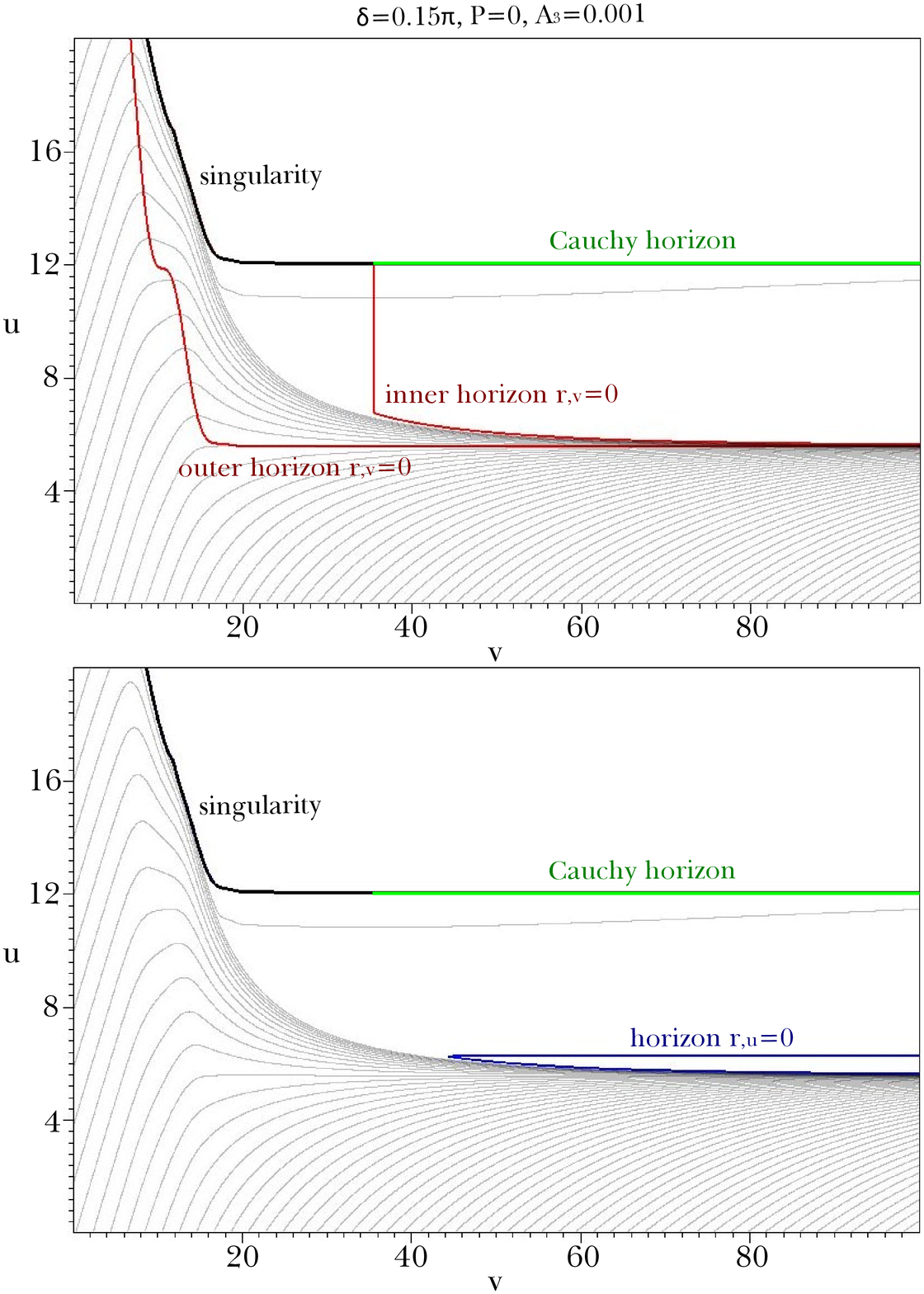}
\caption{\label{fig:3_Ex0.001}Contour diagrams of radial function $r$ and horizon structures for $A_{3}=0.001$. As the exotic matter field is supplied from $v=20$, the outer horizon becomes time--like, and both $r_{,u} = 0$ and $r_{,v} = 0$ types of inner horizons arise. The resulting structure is similar to the one in Figure \ref{fig:d0.15r}, where the semiclassical back--reaction is used.}
\end{center}
\end{figure}

\begin{figure}
\begin{center}
\includegraphics[scale=0.3]{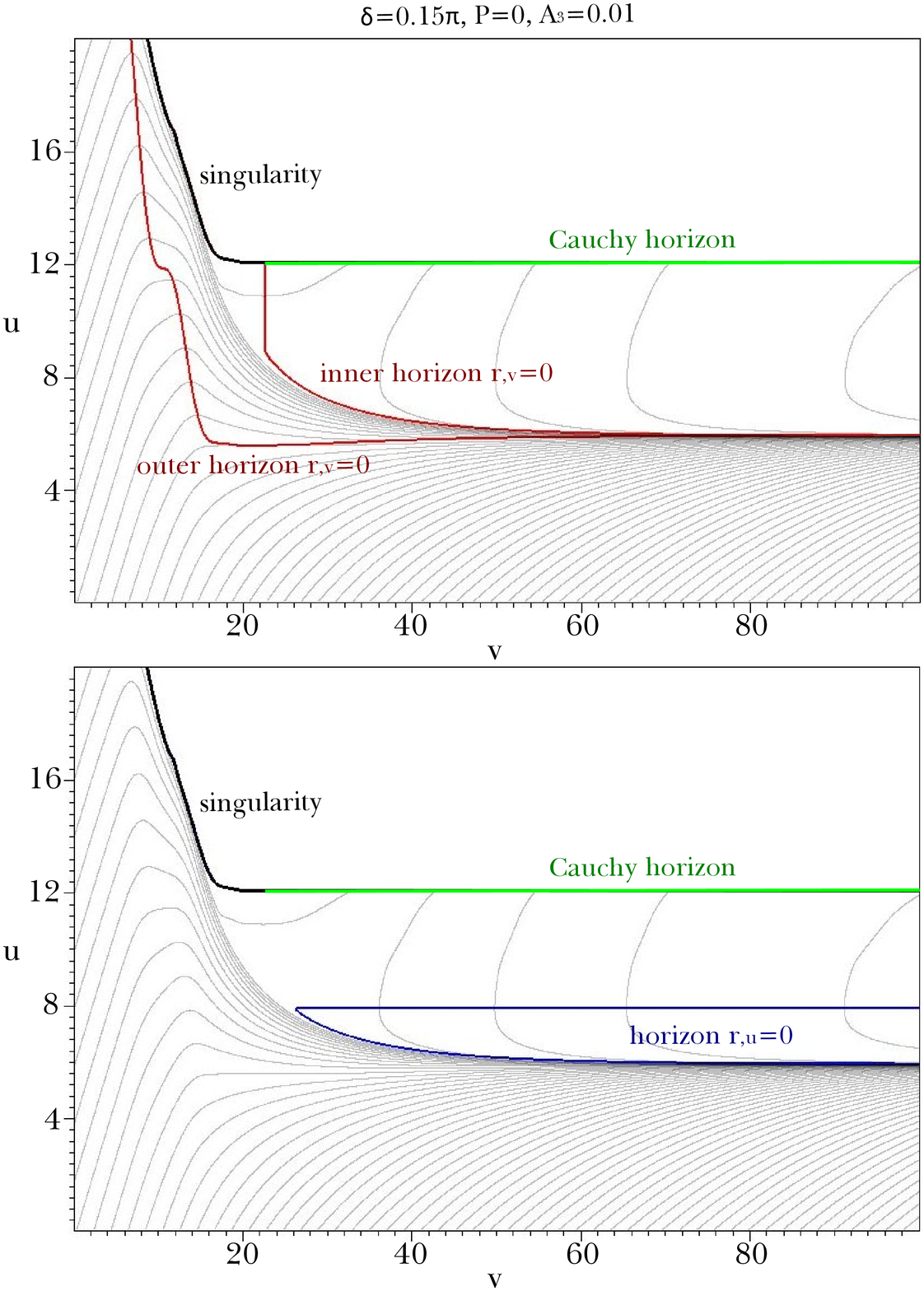}
\caption{\label{fig:3_Ex0.01}Contour diagrams of radial function $r$ and horizon structures for $A_{3}=0.01$. As a larger amount of negative energy is supplied, the $r_{,u} > 0$ region inside of the black hole is increased. The relation between the negative energy and the wormhole--like structure is discussed in Section \ref{sec:Wormhole}. The outer horizon approaches to the inner horizon more quickly.}
\end{center}
\end{figure}

\begin{figure}
\begin{center}
\includegraphics[scale=0.3]{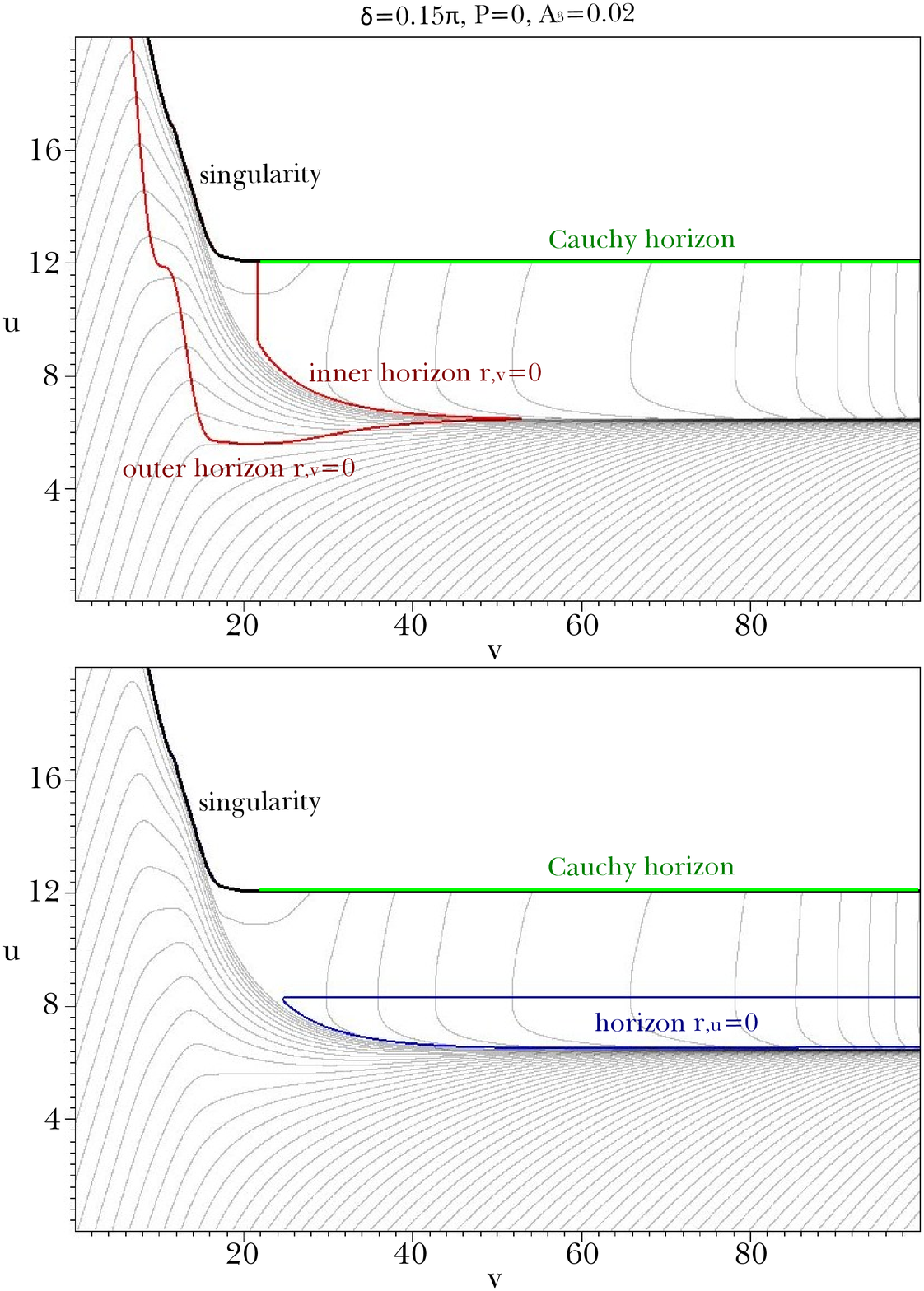}
\caption{\label{fig:3_Ex0.02}Contour diagrams of radial function $r$ and horizon structures for $A_{3}=0.02$. As $Q/M$ reaches unity, the outer horizon and the $r_{,v} = 0$ type inner horizon meet together and disappear, which is consistent with the RN metric in Figure \ref{fig:charged_metric} (c). On the other hand, the $r_{,u} = 0$ type inner horizon maintains its structure.}
\end{center}
\end{figure}

\begin{figure}
\begin{center}
\includegraphics[scale=0.7]{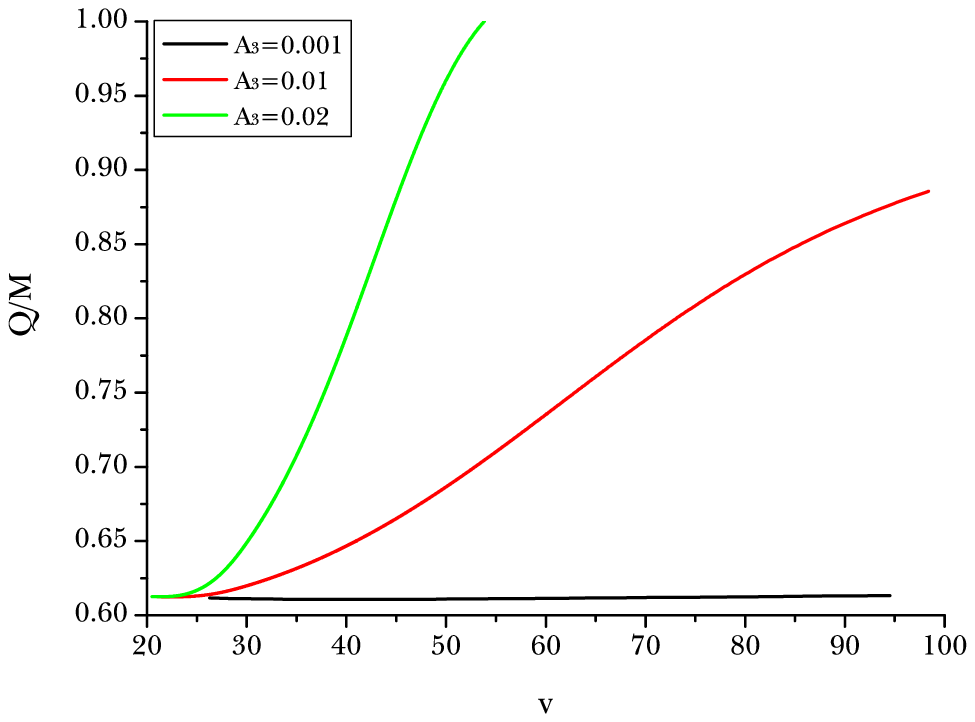}
\caption{\label{fig:QM_exotic}$Q/M$ measured at the outer horizon along the advanced time $v$.}
\end{center}
\end{figure}

Figure \ref{fig:3_Ex0.001}, Figure \ref{fig:3_Ex0.01}, and Figure \ref{fig:3_Ex0.02} describe the results using $A_{3}=0.001$, $A_{3}=0.01$, and $A_{3}=0.02$, respectively.
In all diagrams, time--like outer horizons and space--like inner horizons are observed, which is similar to Figure \ref{fig:d0.15r}.
It supports the idea of using an exotic matter field in studying the long--term evaporation because it affects the horizon structure in a similar way with the semiclassical back--reaction even though they are fundamentally different.
Figure \ref{fig:QM_exotic} shows the $Q/M$ of each black hole measured along the advanced time.
As $A_{3}$ is increased, a larger amount of negative energy quickly reduces $M$.
With $A_{3}=0.02$, $Q/M$ reaches unity in the scope of simulation.
In this case, the time--like outer horizon and the space--like $r_{,v} = 0$ type inner horizon meet together and disappear, which corresponds to Figure \ref{fig:charged_metric} (c).
Note that $Q/M$ cannot exceed unity effectively via Hawking radiation.

The results above provide some implications on the long--term evaporation.
Firstly, larger $r_{,u} > 0$ region is observed in simulations with larger $A_{3}$.
It is consistent with the results in Section \ref{sec:InsideStructure}, where we observed larger $r_{,u} > 0$ region in black holes with smaller $Q/M$ because their high Hawking temperature can be matched to the large negative energy density.
Secondly, the qualitative behavior of the horizon structure is not changed during the substantial mass reduction.
This is in contrast with Section \ref{sec:InsideStructure}, where black holes with larger $Q/M$ have a smaller $r_{,u} > 0$ region and no inner horizons are observed in nearly extreme black holes.
In further simulations with various configurations of exotic matter field, we found that the internal structure becomes insensitive to the change of $Q/M$ or the negative energy density with the lapse of time.

To sum up, $Q/M$ or negative energy density determines the internal structure in the stage of formation.
However, it becomes insensitive to them with the lapse of time, and the structure is fixed.
The reason seems to come from mass inflation.
It induces large curvature to the internal structure, which makes it insensitive to the late--time perturbations.
Eventually, the curvature will become trans--Planckian.
In this regime, the inner horizon can be regarded as a physical singularity with a finite radius, which is analogous to some dilaton black holes models \cite{Hong:2008mw}\cite{Garfinkle:1990qj}.

\begin{figure}
\begin{center}
\includegraphics[scale=0.6]{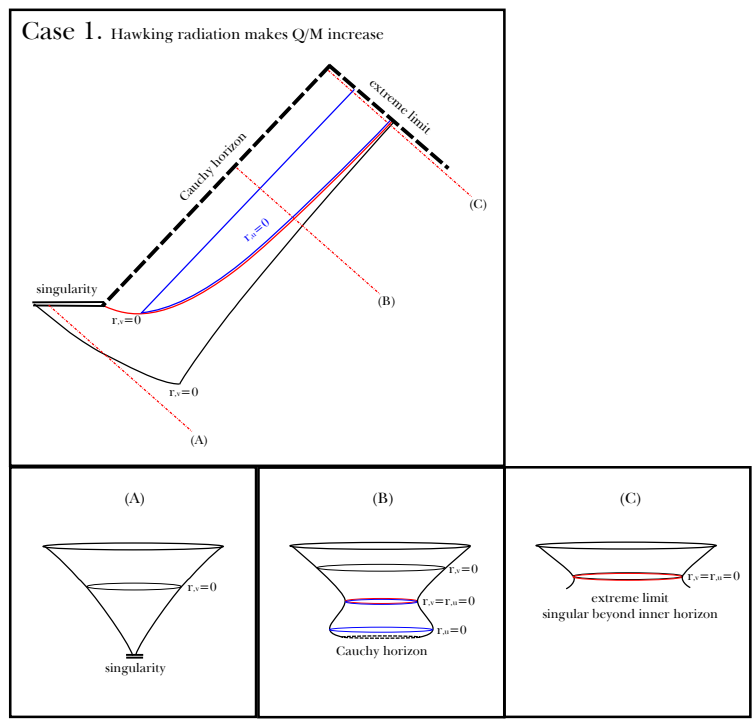}
\caption{\label{fig:diagram_case1}Long--term evaporation of charged black holes. Diagrams in the lower row show schematic diagrams of the area measured by an in--going null observer along the red dotted lines in the upper diagram.}
\end{center}
\end{figure}

We summarize the results in this section in Figure \ref{fig:diagram_case1}.

\section{\label{sec:Neutralization}Neutralization of charged black holes}
As a charged black hole gets some opposite charge, it discharges and becomes a neutral black hole.
In this section, we investigate the neutralization of charged black holes for both non--evaporating and evaporating cases.
For all simulations, we set $A_{2}=0.055$, $\delta=0.15\pi$, and $P=0$.

\begin{figure}
\begin{center}
\includegraphics[scale=0.3]{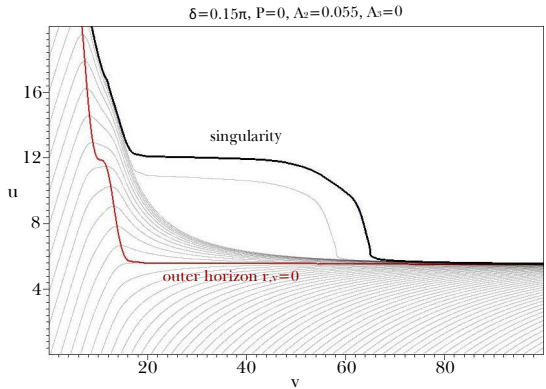}
\caption{\label{fig:3_NoHR_Neut_f}Contour diagram of radial function $r$ and horizon structure for $A_{3}=0$. The charged black hole in Figure \ref{fig:d0.15P0r} discharges and becomes a neutral black hole.}
\end{center}
\end{figure}

We first investigate the neutralization of non--evaporating charged black holes, described in Figure \ref{fig:3_NoHR_Neut_f}.
Compare this result with Figure \ref{fig:d0.15P0r}, which describes the same initial black hole, but not the neutralized one.
As the black hole discharges, the radial function beyond the outer horizon decreases, and the singularity moves downward.
As a result, the gap between the singularity and the outer horizon gradually disappears, and the structure becomes a neutral one.
An out--going null observer who would reach the inner--Cauchy horizon in Figure \ref{fig:d0.15P0r} will hit the central singularity if the black hole is neutralized.

\begin{figure}
\begin{center}
\includegraphics[scale=0.3]{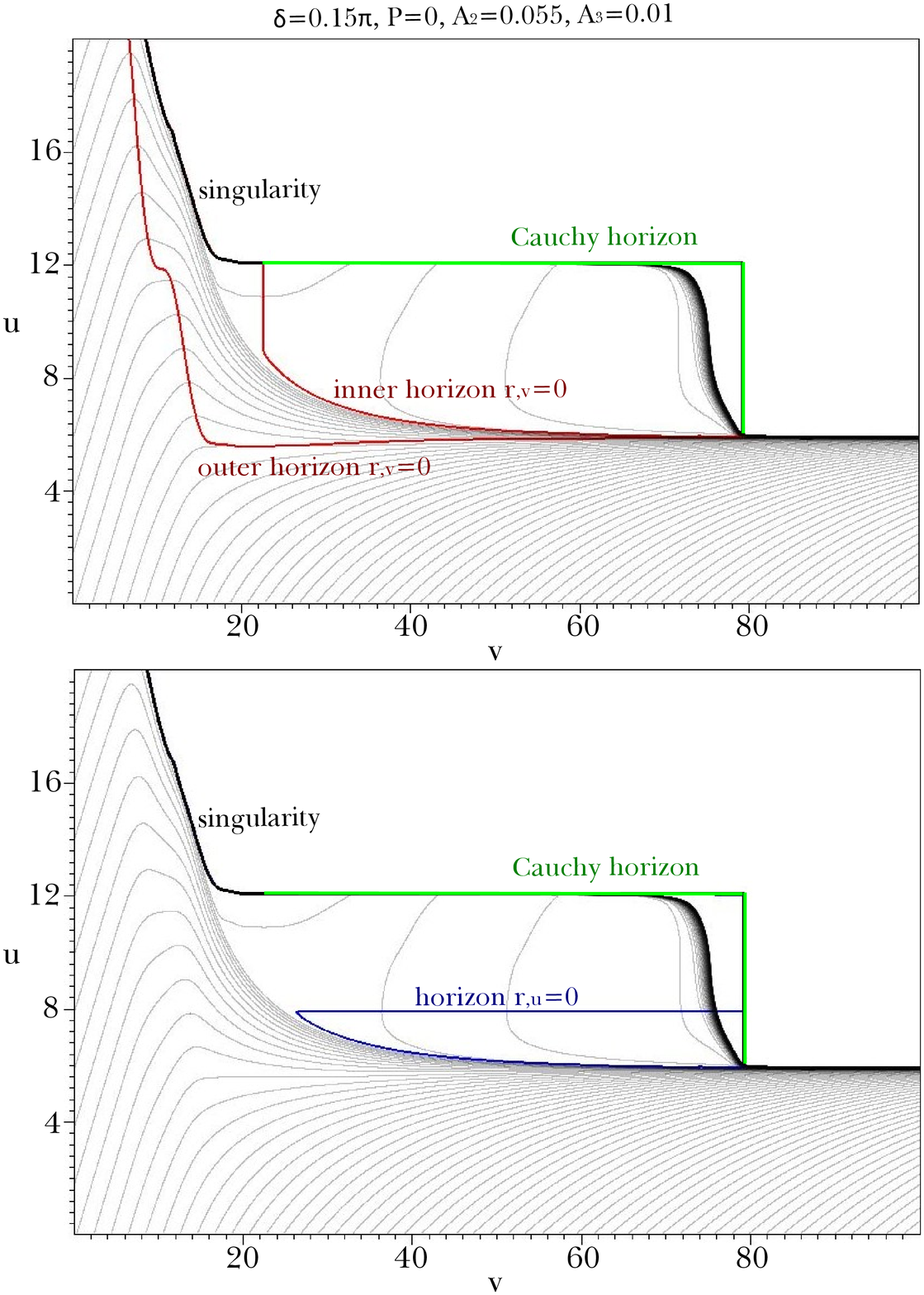}
\caption{\label{fig:3_Ex001_Neut}Contour diagrams of radial function $r$ and horizon structures for $A_{3}=0.01$. An exotic matter field is added to the discharging black hole in Figure \ref{fig:3_NoHR_Neut_f} to consider Hawking radiation.}
\end{center}
\end{figure}

In the previous paper, the neutralization of evaporating charged black holes is studied using the S--wave approximation of semiclassical back--reaction \cite{Hong:2008mw}.
The key feature of this process is the inner horizon evolving to a singularity.
However, the S--wave approximation causes an artificial singularity that disturbs a direct observation of the formation of the singularity.
To avoid this problem, we adopt an exotic matter field to produce an evaporating internal structure and re--test the neutralization process.
We add an exotic matter field with $A_{3}=0.01$ to the discharging black hole in Figure \ref{fig:3_NoHR_Neut_f}.
The result is displayed in Figure \ref{fig:3_Ex001_Neut}.
Similar to the non--evaporating case, the radial function beyond the outer horizon decreases as the black hole is neutralized.
However, the local minimum of radial function for a given $v$ is the $r_{,v} = 0$ type inner horizon in evaporating charged black holes.
Therefore, the decreasing radial function becomes a singularity at the inner horizon, and the region beyond the inner horizon is causally disconnected, leaving the structure of neutral black holes.
This result is the same as the analysis using the semiclassical back--reaction, which confirms the previous results \cite{Hong:2008mw}.

\begin{figure}
\begin{center}
\includegraphics[scale=0.6]{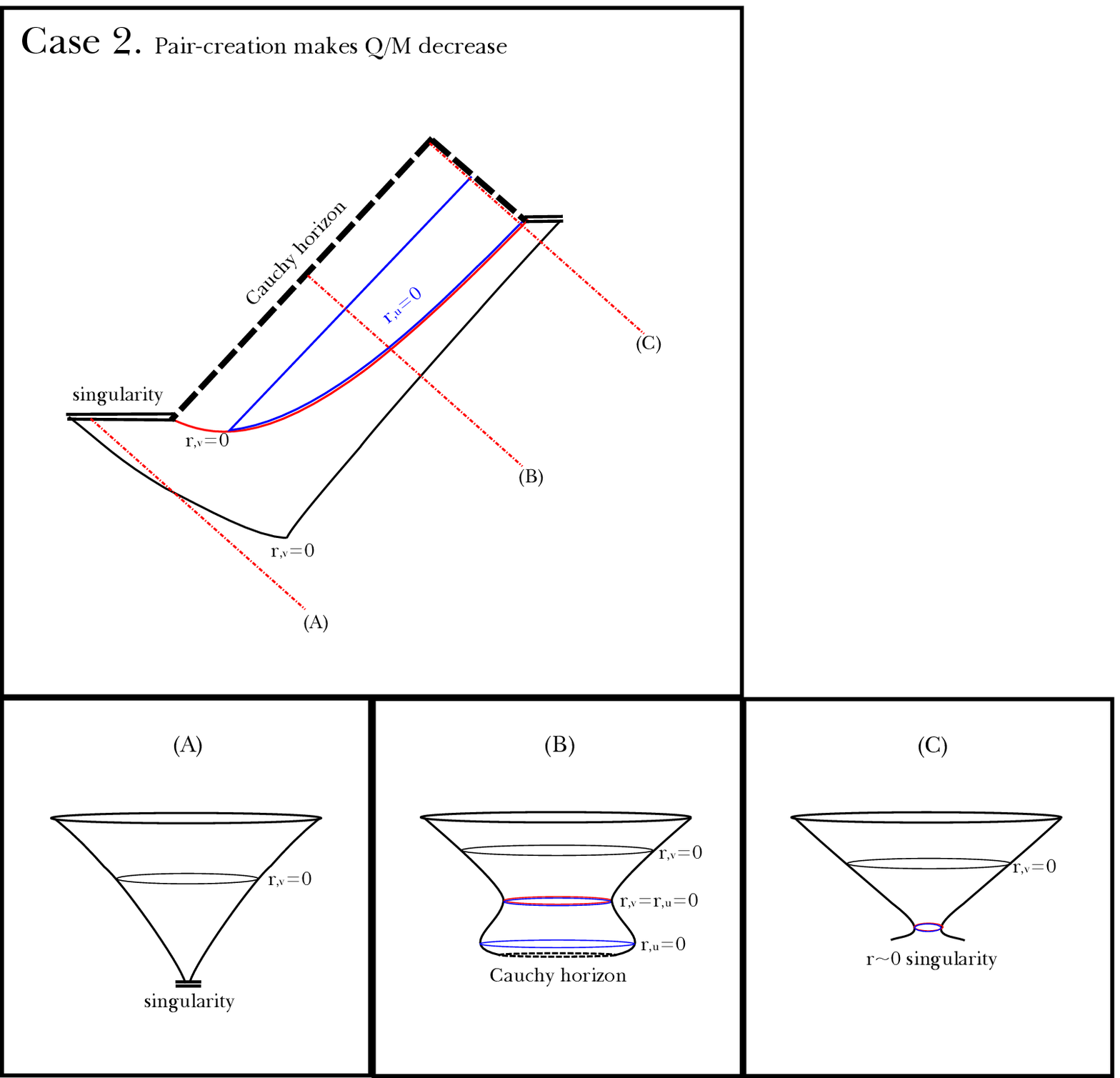}
\caption{\label{fig:diagram_case2}Neutralization of evaporating charged black holes. Diagrams in the lower row show schematic diagrams of the area measured by an in--going null observer along the red dotted lines in the upper diagram.}
\end{center}
\end{figure}

We summarize the neutralization of evaporating charged black holes in Figure \ref{fig:diagram_case2}.

\section{\label{sec:Wormhole}Charged black hole as a wormhole--like structure}
\begin{figure}
\begin{center}
\includegraphics[scale=0.6]{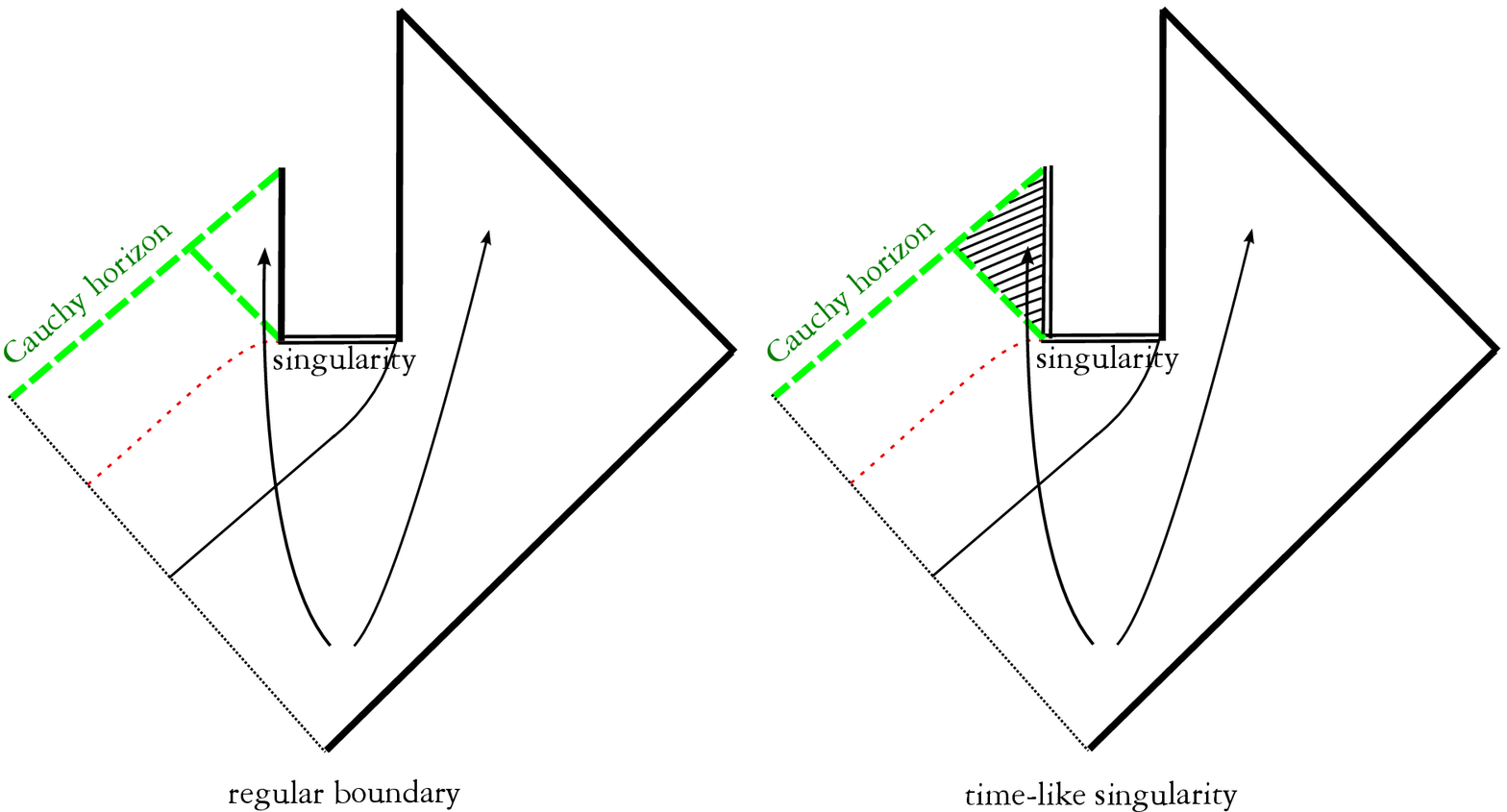}
\caption{\label{fig:wormhole}The neutralization of charged black holes can be identified with the separation between the inside and the outside of the wormhole--like structure. (a) If the inside region is regular and does not have a time--like singularity, this process can cause issues related to the information loss problem. (b) In other cases, semiclassical analysis cannot be applied in this region.}
\end{center}
\end{figure}

The RN metric has wormholes called RN wormholes \cite{RN}\cite{Hawking:1973uf}.
However, they are not observable in dynamically formed charged black holes within the static limit because of the mass inflation effect.
The wormhole--like structure appears again as we consider Hawking radiation.
Evaporating charged black holes with relatively small $Q/M$ have both $r_{,u} = 0$ and $r_{,v} = 0$ types of inner horizons, and any physical observers crossing both of them will observe the increase of locally measured area.
The inner horizon corresponds to the throat of the wormhole--like structure, as described in Figure \ref{fig:diagram_new}.

In Equation (\ref{GAndT}), $G_{uu}$ component of the Einstein tensor gives
\begin{eqnarray}
- \frac{2}{r} \left( r_{,uu}-2 \frac{\alpha_{,u}}{\alpha} r_{,u} \right) = 8 \pi T_{uu},
\end{eqnarray}
where $T_{uu}\equiv T_{uu}^{C}+ \langle \hat{T}_{uu}^{H} \rangle$.
To obtain a wormhole--like structure, $r_{,uu} > 0$ around the $r_{,u}=0$ type horizon is necessary.
It requires $T_{uu} < 0$ in this region, violation of the null energy condition.
Since we assumed Hawking radiation, such violation is naturally obtained.
In contrast to many models of wormholes that involve exotic matters or exotic initial conditions, the wormhole--like structure in evaporating charged black holes can be an interesting and natural model to obtain such a structure in a dynamic process \cite{Morris:1988cz}\cite{Morris:1988tu}\cite{Visser:1989kh}\cite{Visser:1989kg}\cite{Poisson:1995sv}.
However, it is not clear whether a charged black hole is indeed a wormhole or not.
While a wormhole connects two asymptotically flat regions, the existence of such a region inside of a charged black hole cannot be determined by semiclassical analysis because of the Cauchy horizon.
Even if a charged black hole is a wormhole, it is one--way traversable and its deep inside has large curvature which can be trans--Planckian depending on $N$.
Nevertheless, the region around the throat is still reliable in a background of realizable $N$.

As a charged black hole is neutralized, its inner horizon evolves to a singularity.
During this process, the inside and the outside of the wormhole--like structure are separated and the throat becomes a singularity.
It induces a Cauchy horizon, beyond which cannot be determined in terms of the semiclassical physics.
Nevertheless, suppose that the inside region is regular and that there is no time--like singularity, as described in Figure \ref{fig:wormhole} (a).
In this case, this process can cause issues related to the information loss problem because the disconnected region can have an asymptotically flat region \cite{Hong:2008mw}\cite{Hawking:1988ae}\cite{complementarity}\cite{susskind}\cite{HYZ}.
However, in other cases such as Figure \ref{fig:wormhole} (b), the semiclassical analysis does not provide any meaningful remarks.

\section{\label{sec:Conclusion}Conclusion}
We studied the internal structure of charged black holes for important stages in its time evolution using numerical simulations.
We adopted a spherically symmetric charged black hole model that includes the renormalized energy--momentum tensor to consider Hawking radiation.
Solving the equations for various initial conditions, we obtained general behavior of charged black holes.

When charged black holes are formed by collapses of charged matter shells, various internal structures are observed depending on their mass and charge.
For black holes with relatively small charge--mass ratio, both trapping and anti--trapping inner horizons are observed.
The internal structure of such black holes is similar to a wormhole.
As we consider black holes with more charge, the inner horizons gradually move out of sight, and they are not observed in nearly extreme black holes.
Since Hawking temperature is close to zero in nearly extreme black holes, it is natural that their structure becomes analogous to the static limit.

As time passes, the mass and charge of the black hole will evolve via Hawking radiation and discharge.
Although the mass and charge determined the internal structure in the stage of formation, we found that the internal structure becomes insensitive to them with the lapse of time.
The reason seems to come from mass inflation which induces large curvature in the internal structure.
The internal structure determined from the formation seems to be maintained during the evaporation.

We also investigated the neutralization of charged black holes for both non--evaporating and evaporating cases.
As charged black holes obtain some opposite charge, the radial function beyond the outer horizon decreases in both cases, causing the singularity to approach to the outer horizon rapidly in non--evaporating charged black holes.
In evaporating charged black holes, obtaining of opposite charge causes the inner horizon to evolve to a singularity.

Finally, we focused on the wormhole--like structure inside of charged black holes which is a worthwhile object of examination because, without any exotic requirements, Hawking radiation naturally induces such a structure.
From this point of view, the neutralization of charged black holes can be interpreted as the separation between the inside and the outside of the wormhole--like structure.
It may cause some issues related to the information loss problem; however, further study is required.

\section*{Acknowledgments}
The authors would like to thank Ewan D. Stewart, Takahiro Tanaka, and Jakob Hansen for discussions and encouragement.
This work was supported by Korea Research Foundation grants (KRF-313-2007-C00164, KRF-341-2007-C00010) funded by the Korean government (MOEHRD) and BK21.
Also, this work was supported by the National Research Foundation of Korea (NRF) grant funded by the Korean government (MEST) through the Center for Quantum Spacetime (CQUeST) of Sogang University with grant number 2005-0049409.

\appendix
\section{\label{sec:Consistency}Convergence and consistency checks}
\begin{figure}
\begin{center}
\includegraphics[scale=0.7]{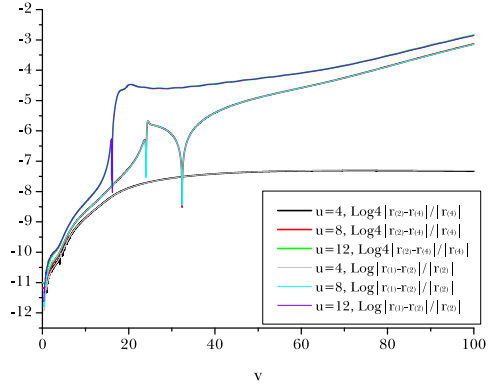}
\caption{\label{fig:convergence}$|r_{(1)}-r_{(2)}|/r_{(2)}$ and $4|r_{(2)}-r_{(4)}|/r_{(4)}$ along a few constant $u$ lines, where $r_{(n)}$ is the radial function calculated in $n \times n$ times finer grid than $r_{(1)}$. It converges to the second order with errors $\lesssim 0.1\%$.}
\end{center}
\end{figure}

\begin{figure}
\begin{center}
\includegraphics[scale=0.7]{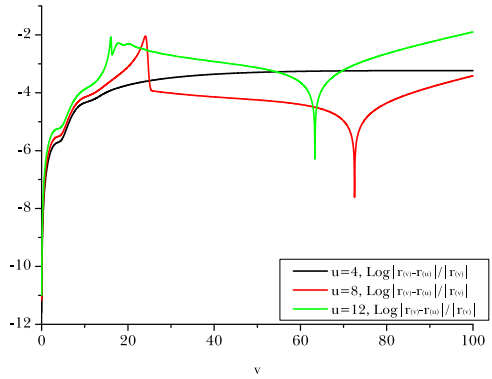}
\caption{\label{fig:consistency}Differences in radial function $r$ between two integration schemes along a few constant $u$ lines. $r_{(v)}$ is calculated by integrating $g$ using Equation (\ref{E4}), and $r_{(u)}$ is calculated by integrating $f$ using Equation (\ref{E2}). The error is $\lesssim 1 \%$.}
\end{center}
\end{figure}

In this appendix, we check the convergence and consistency of the numerical calculations.
The initial condition described in Figure \ref{fig:d0.15r} was tested in this section.
We also successfully tested other initial conditions, which include non--zero $A_{2}$ and $A_{3}$.

To check the convergence, we compared the results from different step sizes: 1 times, 2 times, and 4 times finer for each axis.
Figure \ref{fig:convergence} shows that the differences in radial function $r$ decrease by $1/4$ as the step size becomes half.
The differences converge to the second order with errors $\lesssim 0.1\%$.

To check the consistency, we compared the results from two different sets of equations.
In most of the simulations, we used Equations (\ref{E3}) and (\ref{E4}) to evolve $f$ and $g$, respectively, and $d$, $g$, and $z$ to evolve $\alpha$, $r$, and $s$, respectively.
We compare this result with the other one using Equations (\ref{E2}) and (\ref{E3}) to evolve $f$ and $g$, respectively, and $h$, $f$, and $w$ to evolve $\alpha$, $r$, and $s$, respectively.
The radial function $r$ integrated from each scheme corresponds to $r_{(v)}$ and $r_{(u)}$ in Figure \ref{fig:consistency}, respectively.
The difference is $\lesssim 1 \%$.

Near the horizons and the singularity, the errors increase rapidly, and it comes from the rapidly varying radial function in these regions.
We adopted the adaptive grid refinement method (AGM) to handle this problem \cite{OrenPiran}.
In this paper, we adaptively set the step size $\Delta u$ in $u$ direction so that $r_{,u}\Delta u / r \leq 1 \%$.
When the radial function varies rapidly, AGM automatically reduces the step size so that more detailed calculations can be done in that region.

\end{document}